\documentclass[runningheads,a4paper]{llncs}
\usepackage{amssymb}
\usepackage{algorithm, algorithmic}
\usepackage{graphicx,epsfig}
\usepackage[section]{placeins}
\usepackage[numbers]{natbib}

\usepackage{url}
\urldef{\mailsa}\path|{alfred.hofmann, ursula.barth, ingrid.haas, frank.holzwarth,|
\urldef{\mailsb}\path|anna.kramer, leonie.kunz, christine.reiss, nicole.sator,|
\urldef{\mailsc}\path|erika.siebert-cole, peter.strasser, lncs}@springer.com|    
\newcommand{\keywords}[1]{\par\addvspace\baselineskip
\noindent\keywordname\enspace\ignorespaces#1}
\usepackage{lipsum,graphicx,subcaption}
\usepackage{subcaption}
\usepackage{amsmath}
\begin{document}

\mainmatter  

\title{Immunization Strategies Based on the Overlapping Nodes in Networks with Community Structure}


%
%
\author{Debayan Chakraborty\inst{1} \and Anurag Singh \inst{1} \and Hocine Cherifi\inst{2}}
\authorrunning{Effect of Immunization:Complex Networks}

\institute{Depatment of Computer Science \& Engineering, \\National Institute of Technology, Delhi, New Delhi, India-110040\\
Email: debayan.chakraborty@nitdelhi.ac.in, anuragsg@nitdelhi.ac.in\\
\and
University of Burgundy, LE21 UMR CNRS 6306\\Dijon, France\\
Email: hocine.cherifi@gmail.com
}

%
%

\maketitle

\begin{abstract}
 Understanding how the network topology affects the spread of an epidemic is a main concern in order to develop efficient immunization strategies. While there is a great deal of work dealing with the macroscopic topological properties of the networks, few studies have been devoted to the influence of the community structure. Furthermore, while in many real-world networks communities may overlap, in these studies non-overlapping community structures are considered. In order to gain insight about the influence of the overlapping nodes in the epidemic process we conduct an empirical evaluation of basic deterministic immunization strategies based on the overlapping nodes.Using the classical SIR model on a real-world network with ground truth overlapping community structure we analyse how immunization based on the membership number of overlapping nodes (which is the number of communities the node belongs to) affect the largest connected component size. Comparison with random immunization strategies designed for networks with non-overlapping community structure show that overlapping nodes play a major role in the epidemic process.
\vspace*{-.4cm}
\end{abstract}
\vspace*{-.4cm}
\section{Introduction}
The effect of network structure on the spread of diseases is a widely studied topic, and much research has gone into this field \cite{01,02,03,04,05,06,07,08}. The topological feature of network have been used for immunization within network \cite{pastor2002immunization, gallos2007improving, tanaka2014random, glasser2010evaluation, madar2004immunization, christakis2010social, krieger2003focus}. These works have mainly studied the various immunization strategies and their effect on epidemic outbreak within a social network or contact network.
The study of networks according to the degree distribution, and further the influence of immunization on degree distribution and targeted attacks has been explored by scholars in recent past \cite{cohen2001breakdown, callaway2000network, albert2000attack}. But the community-based study of the network has not received much attention. In this level of abstraction, which has been termed as the mesoscopic level, the concern lies with the properties of the communities. Communities are sets of nodes which show more level of interconnectivity amongst themselves, than with the rest of the network. We can distinguish two type of community structure in the literature depending on the fact that they share some nodes or not. Non-overlapping communities are stand-alone groups where a node belongs to a single community while in overlapping communities a node can belongs to more than one community.  Recent research and analysis of real-world networks have revealed that a significant portion of nodes lies within the overlapping region of two communities \cite{09}. Thus, we look to explore the effect of immunization with these overlapping area of communities on the overall spread of the epidemic within the system. 
Recently the studies of few researchers have considered community structure in the field of epidemiology or pharmaco-vigilance\cite{zhang2013stochastic, 10, becker1998effect}. But, mostly  they have taken these sets as stand-alone groups and have again, not explored the communities as they truly are in real world, overlapping sets of shared nodes. Results in the recent literature show that the knowledge of degree distribution and after that, degree distribution based immunization strategies are not sufficient enough to predict viral outbreak or epidemics in general. Further, the behavior shown by an epidemic on networks with varying community structures also show a certain degree of independence amongst themselves\cite{shang2015epidemic, chen2012epidemic, shang2014overlapping}. Thus, confirming the fact that community structures also play a vital role in the spreading process for epidemics within the network. So community structure has to be factored into the immunization process. In this level of abstraction, the focus lies on nodes of connectivity within two or more communities. In fact, Salathe \textit{et al.} \cite{10} had studied the effect of immunization through these bridge nodes and edges in their paper. However, their community bridge finder model analysed the communities as non-overlapping groups. Further studies have been done by Samukhin \textit{et al.} \cite{12}, who analyzed the Laplacian operator of an uncorrelated random network and diffusion process on the networks. Naveen Gupta \textit{et al.} \cite{11,11b}analyse the properties of communities and the effect of their immunization within their paper. They take the community nodes and analyse them on their out-degree, in-degree and difference of two on the communities to which they belong. Their study shows that community-based degree parameter can help in identifying key structural bridge/ hub nodes within any given network. Their analysis further consolidated the importance of communities and their effect on the overall immunization strategy once they are taken into account. 
The major drawback of all these studies are that; they take networks with no underlying overlap within community structures. Even if there exists a certain amount of overlapping within these networks, they overlook those regions and analyse these areas as independent sets.
In this paper, we look at community overlaps and study their immunization. We analyse the effect of two targeted immunization strategies of nodes within the overlapping regions based on the membership number. We use the classical SIR model of epidemics to analyze the spread of diseases within the network. The Experiment are conducted on a real-world network with ground truth community structure (Pretty Good Privacy). A comparative study with an immunization strategy that is agnostic of the overlapping community structure (random acquaintance \cite{cohen2003efficient}) and an immunization strategy derived for non-overlapping community structure (Community Bridge Finder \cite{10}) is performed.
The remaining of the paper is arranged as follows: 
In Section 2, we present a short introduction to the classical $SIR$ model following which we present briefly the existing immunization strategies which are agnostic to network structure. In Section 3, the overlapping community structure is defined along with the statistic associated with these structures. In section 4 we discuss the experimental results following which we conclude our findings in Section 5. 
 \vspace*{-.4cm}
\section{Background}
\subsection{Classical SIR Model}
The property of the connection of the individual nodes and the nodes that are in the neighborhood have a direct effect on their ability to propagate information within a system, and their ability to stop the information is also worthy.
To characterize the immunization of nodes we first look into the spread of the epidemic within a network.
We present the classical \textbf{SIR} model which we use to study the general characteristic of diffusion within a system.
%
%
%
The model uses rate definitions to define the change of state of each node within the Susceptible, Infected and Recovered states with rates $\alpha$, $\lambda$ and $\beta$.
Whenever infected contacts a susceptible, the susceptible becomes infected at a rate $\alpha$.  
Whenever an infected spontaneously changes to a recovered (simulating the random cure of the individual on diffusion), it does so at the rate $\beta$. $S(t)$, $I(t)$, $R(t)$ gives the evolution of each set within the network. For example, $S(t)$ gives us the fraction of nodes which are susceptible to infection at time $t$. 
The spreading rate $\lambda= \alpha / \beta$ describes the ability of the epidemic to spread within the network. High spreading rate signifies epidemic can spread more quickly within the network.

\vspace*{-.4cm}
\subsection{Immunization strategies}
Largest Connected Component $lcc$ of a network is that component of the network which contains the most number of nodes within them and each node is reachable from every other node. In a sense, no node is dis-connected from another node within the component. In effect, the size of the largest connected component will tell us the maximum limit to which an epidemic can spread. Starting from the full network $\mathcal{N}$ one can remove the nodes according to an immunization strategy and check for the effect on $lcc$. As one transforms the network $\mathcal{N}$, the size of the $lcc$ is also subject to change. The transformed network $\mathcal{N}'$ has the largest component of size $lcc'$. 
With $N$ being the number of nodes in $\mathcal{N}$ and $N'$ being the number of nodes in 
$\mathcal{N'}$, one can say that, since $N > N'$,  $lcc > lcc'$ , $N \in$ $\mathcal{N}$ and $N'$ $\in$ $\mathcal{N}'$. Thus with each transformation of network $\mathcal{N}$ one aims at reducing the size of $lcc$ as much as possible. A good immunization strategy is one, which,  with the least number of nodes removed, transforms the network in such a manner, that the $lcc'$ of the transformed network
$\mathcal{N}'$ is the least. Here, we present stochastic strategies of immunization. Stochastic models are usually agnostic about the global structure and thus are used here for comparative analysis with the proposed strategy which too uses no prior information about the network.
\vspace*{-.4cm} 
\subsection{Random Acquaintance}
Random Acquaintance is one of the stochastic strategies for immunization present in current literature. 
Random Acquaintance was first introduced by Cohen \textit{et al.} in their paper \cite{cohen2003efficient}. It works by picking a random node and then choosing its neighbour at random. A number $n$ is taken before the start of the process and if an acquaintance of the randomly selected node is selected more than or equal to $n$ times, then, it is Immunized. In case where $n=1$, any acquaintance will be immunized immediately. Without any prior global knowledge of the network, this strategy identifies highly connected individuals.
\vspace*{-.4cm}
\subsection{Community Bridge Finder $CBF$}
$CBF$ is a random walk based method to find nodes connecting multiple communities. It was presented first by Salathe \textit{et al.} in their work \cite{10}. A random node is selected at the start, and then a random path is followed until a node is reached which is not connected to more than one of the previously visited nodes. The idea is based on the belief that this node which is not connected to more than one of the previously visited nodes, in the random walk is more likely to belong to a different community. This strategy too has no prior information about the network structure.
\vspace*{-.4cm}
\section{Overlapping Community Structure}
Studies have been carried out by scholars for detecting communities within a network. However, till now, there has been no widely accepted definition of community structures in literature. A common notion is to consider those sets/ groups of nodes which show high inter-connectivity amongst themselves than with the rest of the network, as communities. These densely connected sets of nodes may thus carry structural hubs within themselves. Overlapping communities are sets of nodes with common nodes shared within them. These common nodes of the sets lie within the overlapping region of the communities.The total number of nodes   shared amongst two or more communities, $n_{ov}$ gives us the total number of nodes within the overlapping region in the network.  
\vspace*{-.3cm}
\subsection{Definitions}
\subsubsection{Community size :}
The community size $s$ defines the number of nodes in each community. If $C_1,C_2,C_3....C_z$ signify each of the $z$ communities in a network $\mathcal{N}$ then the size of a community, $s$ is $|C_i|$ for $i \in [1,2,....z]$ and it signifies the number of nodes in the community $i$.
\vspace*{-.4cm}
 \subsubsection{Membership number :}
The Membership, $m$, of each node $i$, $i \in {[1,2,...N]}$ where $N$ signifies the total number of nodes in the network, defines the number of communities to which any node  $i$ belongs to. Thus for  $m$ greater than 1, one can state that the node belongs to the overlapping region in the network. A quick analysis of the degree distribution of nodes within the overlapping regions reveals that the degree distribution of the overlapping region also follows a power law characteristic \cite{09}. 
\\

\vspace*{-.4cm}
\vspace*{-.4cm}
\subsubsection{Membership function :}
The membership function $m()$ gives us the membership number of any node $i$, within the network, provided we know the communities within the network. A finite number of nodes dictate restricted communities present within the network, and thus, the variation of $m$ is also finite. If the largest membership within the network is $x$, the smallest being 0, one can divide the entire network as a group of disjoint sets where each set contains the nodes whose membership number is equal to the membership number of all other elements of the set. Thus, for all $ i  \in  \mathcal{N}$, $m(i)=m$, where $m  \in  [1,2,3....x]$ and $i \in M_m$,   where  $ M_m \in [M_1,M_2,M_3.....M_x]$ and $M_1\cap M_2 \cap M_3 \cap.....\cap M_x =0$. Further, for all $ i,j  \in  \mathcal{N}$
 if  $  i,j  \in M_m$ , then
$m(i)=m(j)=m$. 
\vspace*{-.4cm}
\subsubsection{Overlap size:} The overlap size $s_{ov}$ is the number of nodes shared between any two communities. $C_1,C_2,C_3....C_z$ signify each of the $z$ communities in a network $\mathcal{N}$.
 The intersection of two communities $C_i,C_j$ is given by $C_{ij}$ and the size of the overlapping region $s_{ov}$ is defined as $|C_{i,j}|$ which signifies the number of nodes shared by the two communities.
\vspace*{-.4cm}
\subsection{Immunization of overlapping nodes}
In this work membership based immunization strategy has been proposed. The membership number metric had been explored by Palla \textit{et al.} \cite{09} and we have studied the effect of immunization based on this metric on the overall diffusion process. 
We study the effect of immunization on the $lcc$. We have looked into the importance of high membership nodes as well as low membership nodes. As it is shown in \cite{09} the power law nature of $m$ makes it interesting to analyse the effect of membership number based immunization on the  $lcc$. A strategy based on membership based immunization is proposed here. If nodes $i,j,k,l,......$ are arranged in sequence of their membership number and then removal is initialized, two possible strategies emerge. \\
In our analysis, nodes, $i, j, k, l,...$ are removed  and analysed.\\
For any nodes, $i, j, k,... $,\\
\textbf{Immunization starting from highest overlap membership to lowest overlap membership($HLMI$): }
\begin{center}
$i$ is removed before $j$ and $j$ is removed before $k$ if \\
$\textit{m(i)} >\textit{m(j)}>\textit{m(k)}$, $\forall \textit{i},\textit{j}, \textit{k} \in \mathcal{N}$
\end{center}
\textbf{Immunization starting from lowest overlap membership to highest overlap membership($LHMI$):} 
\begin{center}

$i$ is removed before $j$ and $j$ is removed before $k$ if \\
$\textit{m(i)} < \textit{m(j)} <\textit{m(k)}$, $\forall \textit{i},\textit{j},\textit{k} \in \mathcal{N}$\\
\end{center}
\vspace*{-.4cm} 
	\begin{algorithm}[H]
		\caption{HLMI algorithm}
            \begin{algorithmic}
			
			\STATE \textit{\textbf{Input}} {Graph $G(V,E)$, Membership measure $(m)$, No. of nodes to be immunized}
			\STATE \textit{\textbf{Output}} {Graph with Immunized/Removed Nodes}
            
            \STATE Calculate the Membership number $m$ for  each node in the graph using membership function $m()$ ;
            \STATE  Sort the nodes in decreasing order of their membership values ;
            \STATE  Remove top $n$ nodes with highest membership values from the Graph $G$ ;
            \STATE  Return graph $G$ after removal of  immunized nodes ;
            \end{algorithmic}
         
	\end{algorithm}
\vspace*{-.4cm}   

\begin{algorithm}[H]
	   \caption{LHMI algorithm}
	  \begin{algorithmic}
		
			
\STATE			\textit{\textbf{Input}} {Graph G(V,E), Membership measure $(m)$, No. of nodes to be immunized}
\STATE			\textit{\textbf{Output}} {Graph with Immunized/Removed Nodes}
         \STATE Calculate the Membership number $m$ for  each node in the graph using membership function $m()$ ;
     \STATE Sort the nodes in increasing order of their membership values ;
            \STATE Remove top $n$ nodes with lowest membership values from the Graph $G$ ;
            \STATE Return graph $G$ after removal of  immunized nodes ;
             \end{algorithmic}
             \end{algorithm}
             \vspace*{-.4cm}
The distribution of membership number $m$ and the degree distribution of corresponding set of nodes with equal membership number shows that small membership node sets have a larger number of nodes while large membership node sets have a small number of nodes in them. The outliers of the small membership nodes show that the spread of degree values in each set is vast, and high degree nodes are spread in all the sets of different membership. The small membership number nodes, which are large in number by intuition should perform comparatively better than the immunization removing high membership nodes first since, the small membership set contains more nodes with larger variance in its degree distribution. 
\vspace*{-.4cm}
\section{Experimental Results}
\subsection{Analysis of the PGP dataset}
We have used the Pretty Good Privacy network \cite{hric2014community}dataset for our work. The network consists of email addresses which have signatures associated with them. The groups in this network are the email domains which are present in the dataset as ground-truth communities, where every node explicitly states its full involvement in the community it belongs. The network does show a certain degree of overlap amongst its various groups. The dataset consists of 81036 nodes and 190143 edges, with 17824 groups. Further, the link between two nodes is undirected and un-weighted in nature. The network confirms to the power-law degree distribution followed by scale free networks and contains no cyclic or multiple edges. The ground-truth community structures have been studied and their overall effect on epidemic/ diffusion process has been analysed. Our analysis of ground truth data shows us that the size of communities within a network follow a power- law property confirming with the existing literature \cite{09}. Thus, there exists a significant number of small communities and comparatively less number of communities which are quite large in size.
The underlying overlapping nature of the communities present in the dataset is explicitly shown within the ground-truth community data. We find the nodes within the overlap regions and study the size of all the overlapping areas within the network. It too showcases power-law characteristics to a certain degree. 
The membership number $m$ too follows a power-law characteristic. Hence, a larger portion of the nodes within the network shows lower membership to communities.
The result for $m$ and the distribution for $s_{ov}$ together give us an interesting insight into the nature of overlap within the network and the participating nodes. This distribution confirms with the existing literature \cite{09} wherein we see the inherent power-law nature of the metrics associated with any node within a community. More nodes were found to exist within the lower membership sets while higher membership set showcased less number of nodes present within them. Also, the number of outliers for lower membership nodes are large in number. One may conclude that the high degree nodes are spread through all memberships within the network. It is thus an important reason to study the immunization from two viewpoints as explained in Section 3. 
\begin{figure}[htb!]
\begin{center}
$\begin{array}{cc}
\centering
\includegraphics[width=0.5\linewidth, height=1.98 in]{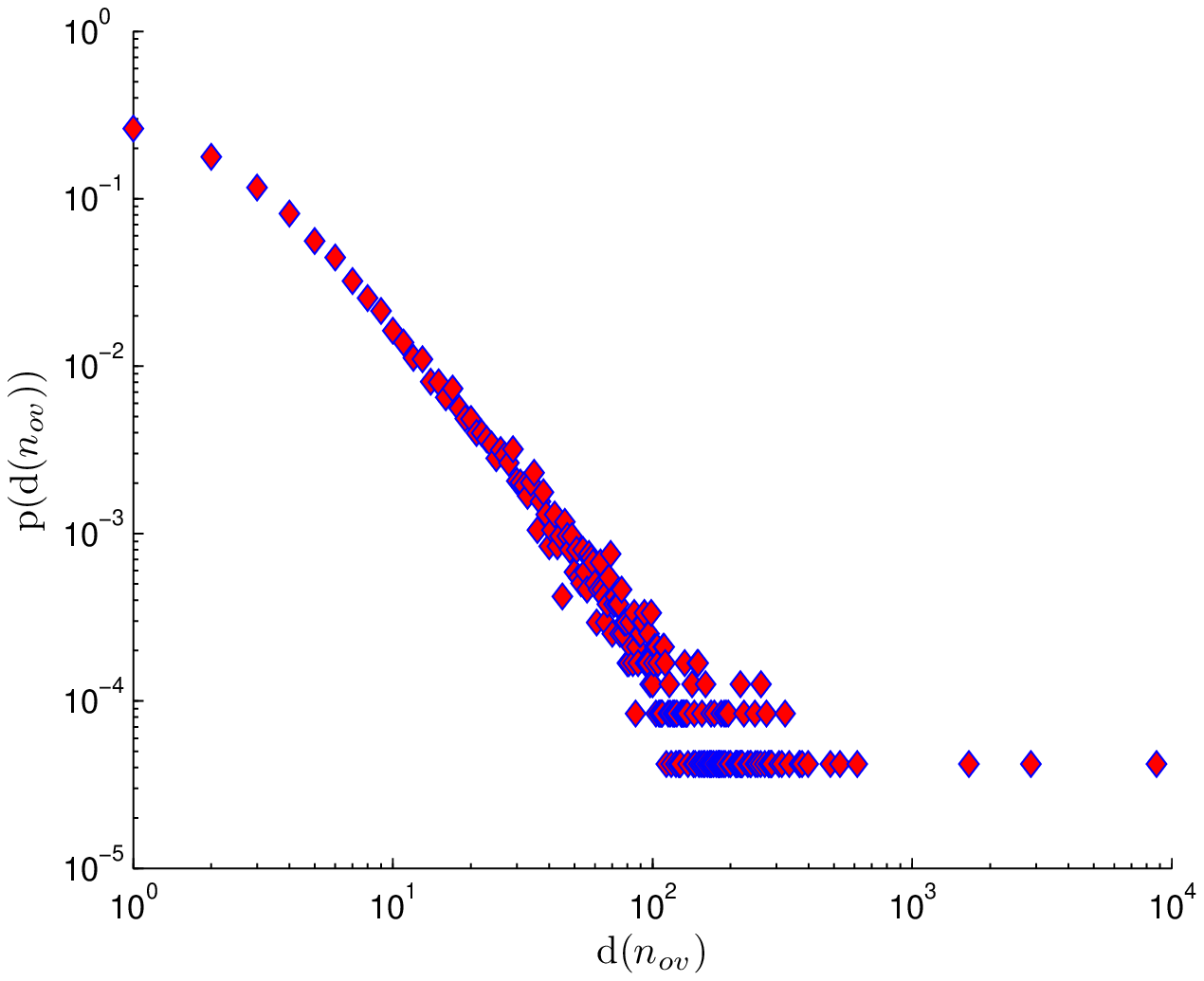} &
\includegraphics[width=0.5\linewidth, height=1.98 in]{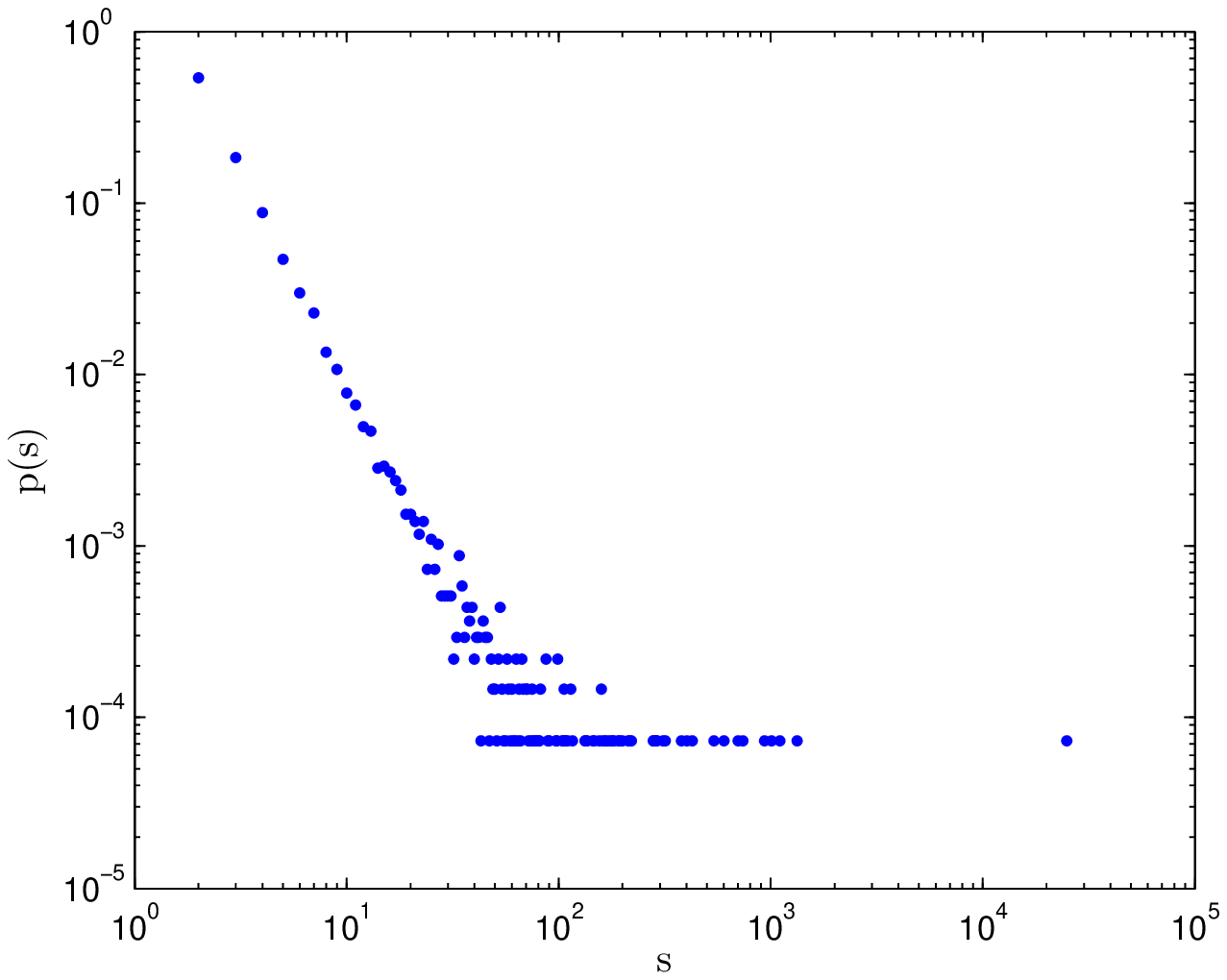} \\
\mbox{(a) } & \mbox{(b)} \\
\includegraphics[width=0.5\linewidth, height=1.98 in]{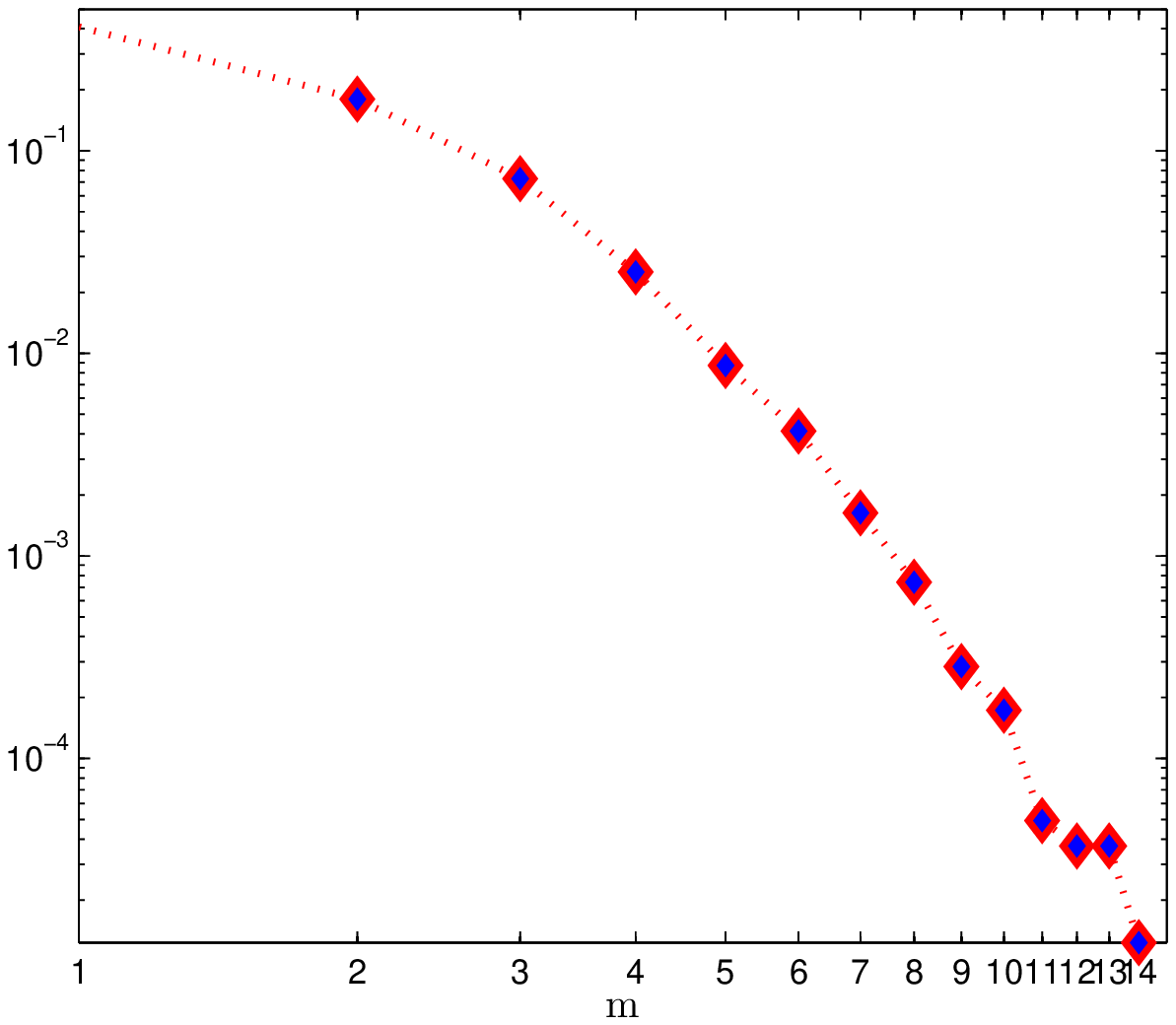} &
\includegraphics[width=0.5\linewidth, height=1.98 in]{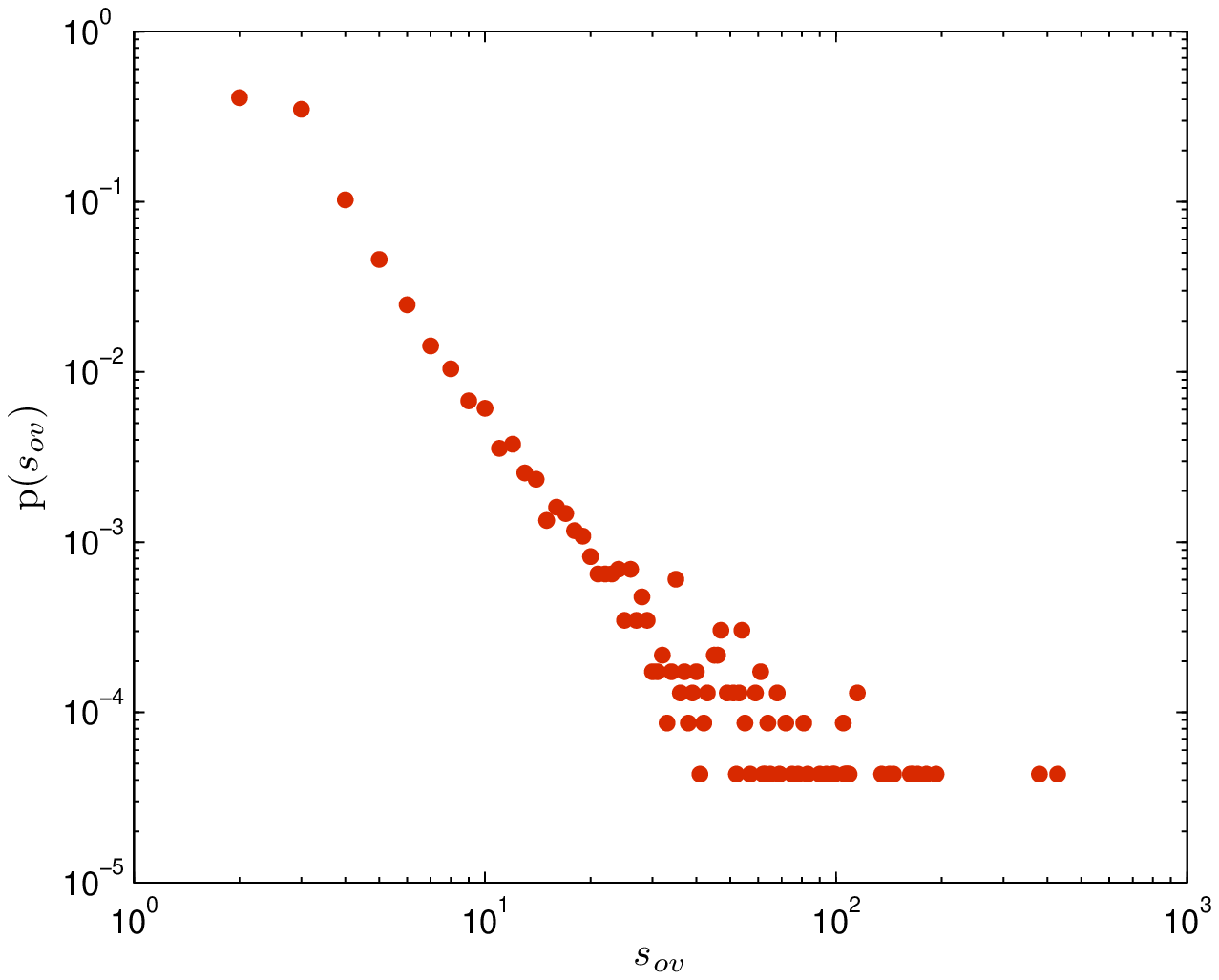} \\
\mbox{(c) } & \mbox{(d) } \\
\end{array}$
\end{center}
\caption{Figure \textbf{(a)} shows the degree distribution in the overlapping nodes within the network.
		 Figure \textbf{(b)} shows the variation of community sizes in the network. Figure \textbf{(c)}the cumulative degree distribution of the membership number $m$. Figure \textbf{(d)} the cumulative distribution of the overlap size. All the above studies were made on the PGP dataset.} \label{f1}
\vspace{-1em}
\end{figure}
 The analysis of community metrics mainly $s$, the size of communities, $s_{ov}$, the size of the overlap region, membership number $m$, and the degree distribution of the nodes in the overlap region $n_{ov}$, are presented in Figure \ref{f1}. We show that all the metrics follow the characteristic as given by Palla \textit{et al.} \cite{09}. The metric $d(n_{ov})$ signifies the degree of the nodes within overlapping region in the network.
 The function $p(.)$ of any metric, signifies the cumulative distribution of the corresponding metric. It might be concluded that the PGP network dataset and the facebook dataset adheres to all the prerequisites for accepting the PGP and facebook network as a network containing overlapping communities within the network.
\vspace*{-.4cm}
\subsection{Evaluation of the Immunization Strategies}
\begin{figure}[htb!]
    \centering
    \includegraphics[height=2.05 in,width=.7\textwidth]{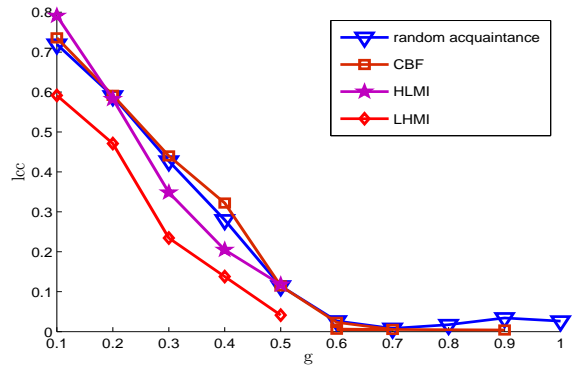}
   \caption{Fraction immunized $(g)$ and its effect on the largest connected component $(lcc)$}
    \label{f2}
\end{figure}

\begin{figure}[htb!]
\centering

  \subcaptionbox{Fig (a)}[.45\textwidth][c]{%
  \includegraphics[width=.49\textwidth, height=1.9 in]{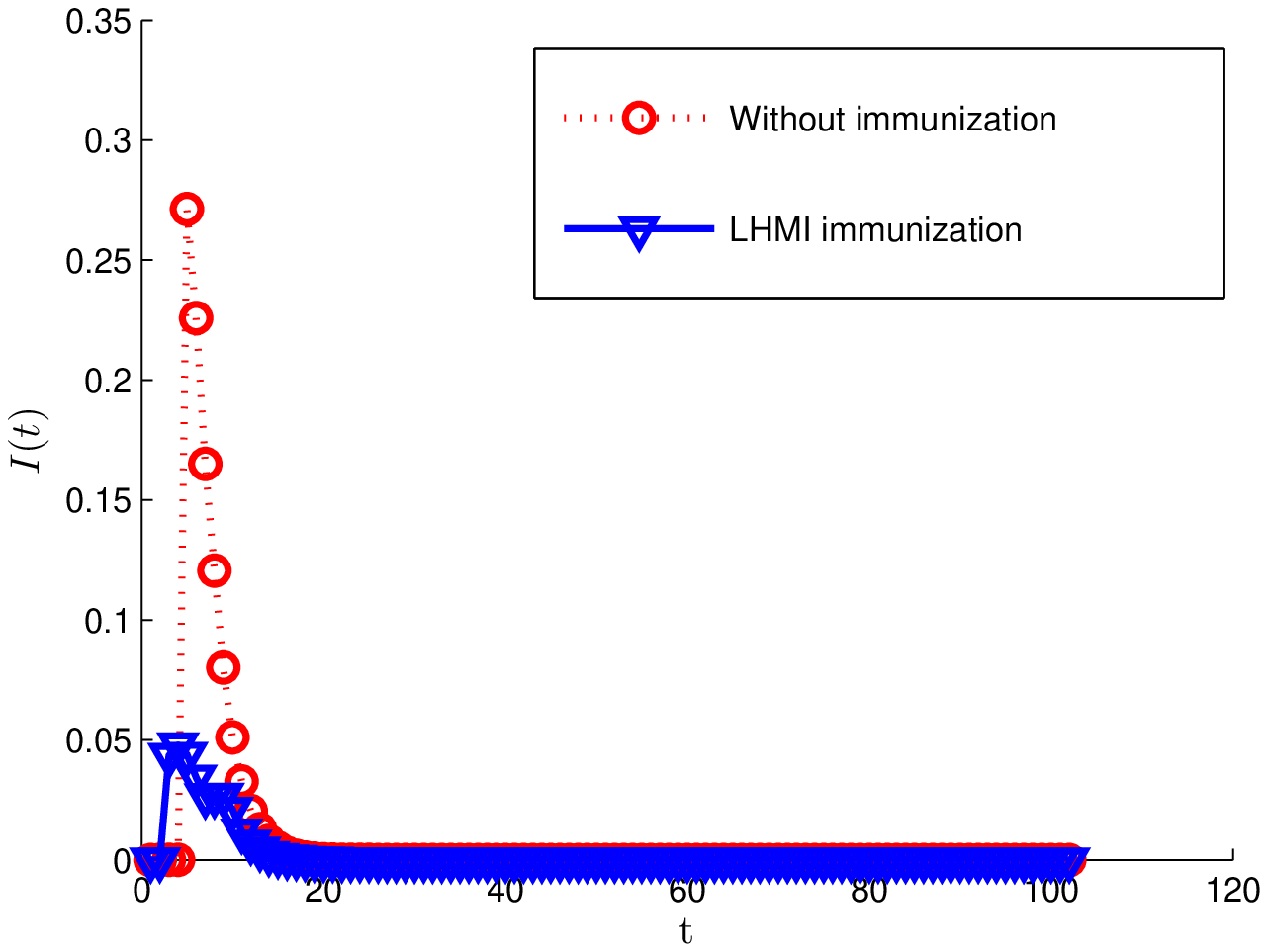}}\qquad
  \subcaptionbox{Fig (b) }[.45\textwidth][c]{%
  \includegraphics[width=.49\textwidth, height=1.9 in]{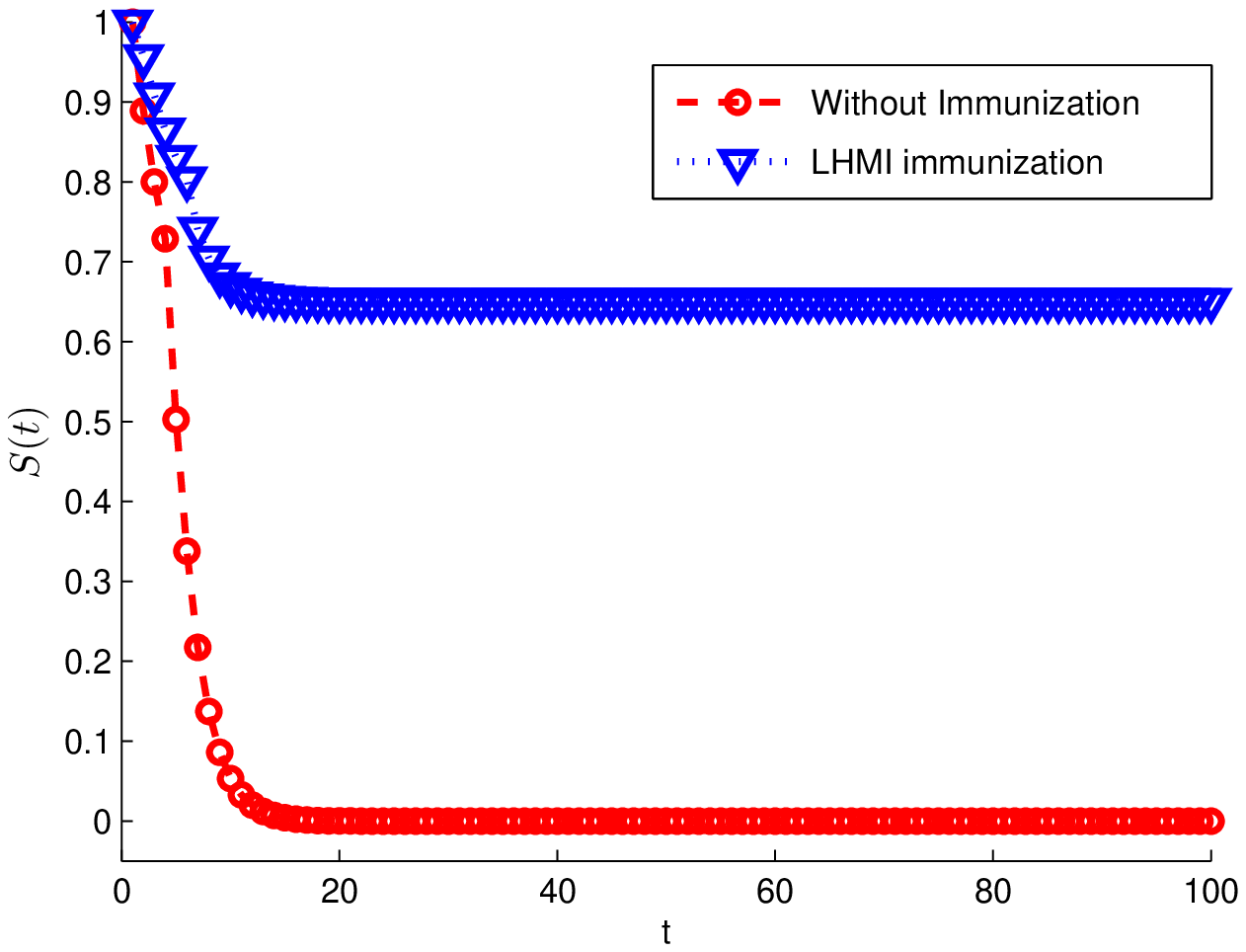}}\qquad
  \subcaptionbox{Fig (c)}[.5\linewidth][c]{%
  \includegraphics[width=.49\textwidth, height=1.9 in]{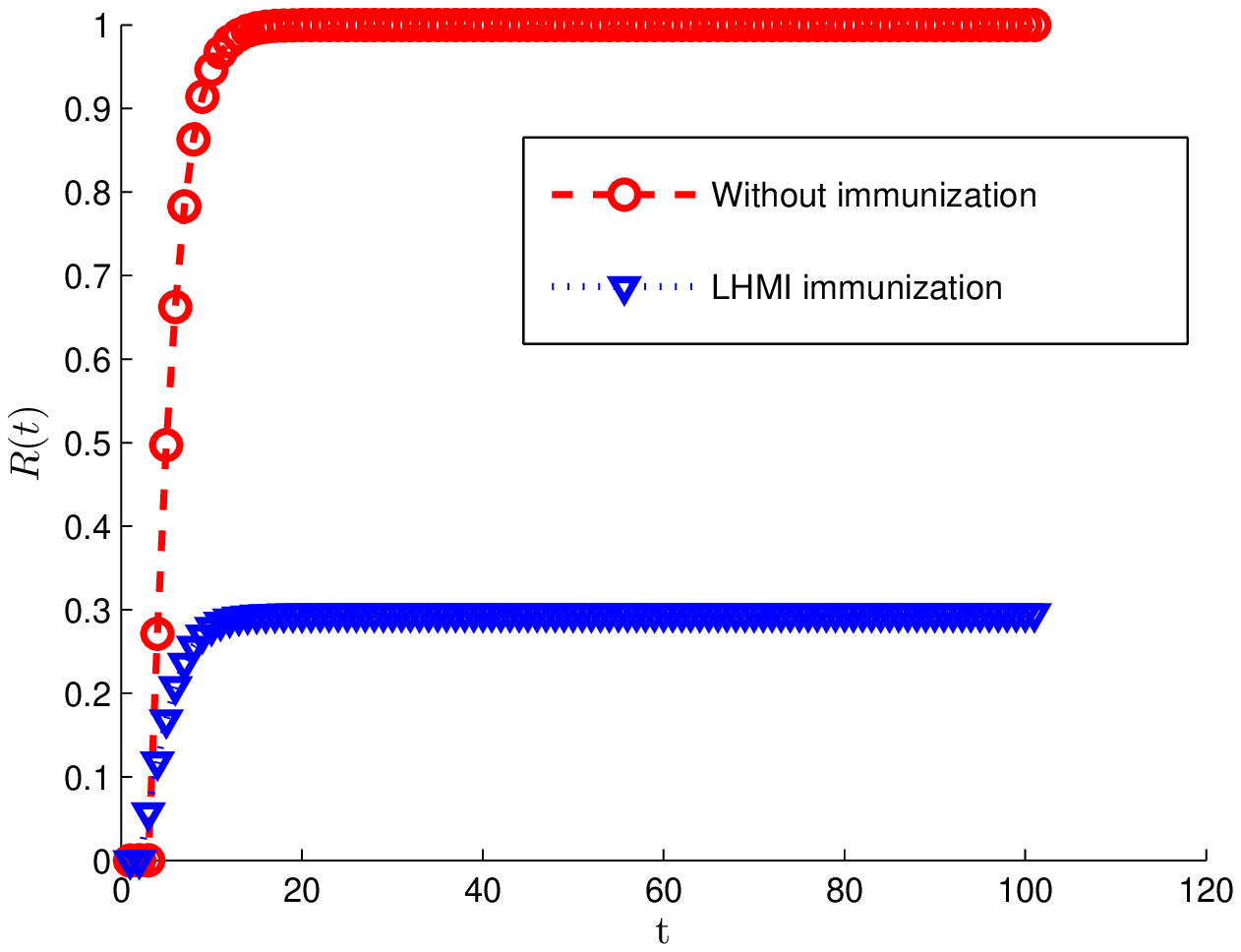}}
\caption{\textbf{(a)} shows the time evolution of the fraction of infected nodes, ${I(t)}$, \textbf{(b)} shows the time evolution of the fraction of susceptible nodes, ${S(t)}$ and \textbf{(c)} shows the time evolution for the fraction of recovered nodes ${R(t)}$ within the network } \label{f3}
\vspace{-1em}
\end{figure}

The comparative results for the $CBF$ and random acquaintance have been studied and have been presented in Figure \ref{f2}. We find that the immunization strategy based on removal of nodes in overlap region show a comparatively better performance than the stochastic methods of immunization. We see that the performance of $CBF$ and random acquaintance are acceptable only after fifty percent of the nodes have been removed from the network. This necessitates that half or more of the entire population be vaccinated/ immunized to ensure no outbreak. Although it is a good approach still the requirement to immunize half of the population becomes a problem. Considering areas of vast population or even highly populated network of nodes, this strategy presents us with an uphill task of immunizing a considerable amount of the population. It is a hefty price to pay, but at the same time it is a must since $CBF$ and random acquaintance being stochastic strategies have no knowledge of the entire network. A similar fraction of nodes is required to be immunized if we have an idea about the communities present within the network and follow the $HLMI$ strategy of immunization. It does not give an overall better performance than the stochastic methods. However, its performance is comparable to the one for $CBF$ and random acquaintance and is thus a strategy which may be accepted as at par with the existing stochastic methods. The $HLMI$ strategy, on the other hand, outperforms stochastic based strategies at lower levels of immunization. One point which becomes important if the community knowledge is readily available. It must be understood that the $HLMI$ strategy gives good performance when half of the population is vaccinated. But, the performance at lower levels of immunization at the same time brings to notice that there is a chance of a trade off between the two groups of strategies. Thus, it gives us a viable alternative approach to solving the same problem, whereas, with varying cases, we may take varying options as and when we feel suited to the need of the hour. Next, we look into the performance of the $LHMI$ strategy. As lower membership nodes are more in number, owing to the power-law characteristic followed by the membership number $m$ \cite{09} it comes as no surprise that we have more variance in the degree distribution of the nodes in the sets of smaller $m$ values. Also, this variance may play an important part and have an overall better effect in the immunization process. From Figure \ref{f2} it is evident that the immunization on the network as a consequence of $LHMI$ outperforms all the other three strategies namely $CBF$, random acquaintance and $HLMI$. At every level of immunization, $LHMI$ gives a performance which $HLMI$  achieves only at the next level. Throughout the levels, it consistently outperforms the stochastic strategies. The nodes which are immunized first in $LHMI$ strategy being part of communities are themselves part of small clusters. Also, since small clusters dominate in any network as shown by Palla \textit{et al.} \cite{09} the $LHMI$ strategy can distort/ break connections in a larger part of the network than the $HLMI$ strategy. The small clusters which populate a network more contain a wider range of degree distribution, and thus, their removal has a better effect on reducing the largest component size. Thus, as is evident from our findings, the $LHMI$ strategy gains an advantage to other strategies in the immunization procedure and hence outperforms stochastic strategies. In both $HLMI$ and $LHMI$, the abrupt halt in the studies of membership based immunization is due to the absence of further nodes belonging to communities within the network. At forty percent removal, the performance of $LHMI$ is worthy. From the results, it is evident that immunization based on $LHMI$ strategy is better at reducing the size of the largest connected component than the one following $HLMI$ strategy. Therefore, one may further conclude that the removal of sets with higher variance in the degree distribution of their nodes give a comparatively better result than those where the number of nodes and variability in their degree distribution is less. 
\vspace*{-.4cm}
\vspace*{-.4cm}
In Figure \ref{f3} the evolution of Infected (I), Susceptible (S) and Recovered (R) nodes are shown.The evolution in the network after immunization based on our proposed strategy ($LHMI$) is compared with the evolution within the network with no immunization. Initially we start with one infected node within the network and we study the gradual evolution at consequent time evolution. The value of $\lambda$, is set to 1. Comparative studies in all these figures show that our proposed method ($LHMI$) is efficient in arresting the fast growth within the system and thus is capable of stopping an epidemic from occurring. The overall performance may be attributed to the stopping of infection spread, which the $LHMI$ algorithm does quite efficiently in the beginning stages of diffusion. Thus our proposed strategy based on membership based immunization is a viable alternative to the stochastic strategies present in the current literature.  
\vspace*{-.4cm}
\vspace*{-.4cm}
\section{Conclusion}
The global topological information of a network is not always available to us.
 Thus, the requirement of procedures which utilize another available information of communities is needed. In the results of our study, we have analysed the effect of local community information(present in ground truth communities) based immunization strategy on real world network of a vast number of nodes. The membership number based calculation is dependent solely on the knowledge of the communities in the network. We see that $LHMI$ and $HLMI$ give results which are comparable to stochastic models of immunization and work on par with the same efficiency if not better. We require no knowledge of the network, and yet the achieved results surpassed the stochastic model performances which need at least some local connection information of the studied nodes. Thus, we find that community information may be effectively utilized for developing efficient immunization strategies. 
\bibliographystyle{splncs03}
\bibliography{references}

\end{document}